\documentstyle[11pt,mrs2003,epsfig]{article}
\newlength{\textwidthhalf}
\newlength{\textwidthrhalf}
\newcommand{\average}[1]{\left\langle #1 \right\rangle}

\newcommand{\Cov}{\mbox{Cov}}

\newcommand{\dd}{{\mathrm d}}

\newcommand{\Map}{{M_{\rm ap}}}

\newcommand{\tot}{{\rm tot}}

\newcommand{\Omegam}{\Omega_{\rm m}}

\newcommand{\ee}{{\rm e}}

\newcommand{\Mapsq}{M_{\rm ap}^2}

\begin{document} 
%


\title{COSMOLOGICAL PARAMETERS FROM COSMIC SHEAR}

\author{MARTIN KILBINGER}
\affil{Institut f\"ur Astrophysik, Universit\"at Bonn}

\begin{abstract}
We present simulations of a cosmic shear survey and show how the survey geometry
influences the accuracy of determination of cosmological parameters. We
numerically calculate the full covariance matrices of the two-point statistics
$\xi_+$, $\xi_-$ and $\average{\Mapsq}$ and
use maximum likelihood and Fisher matrix analyses in order to derive expected
constraints on the cosmological parameters $\Omegam, \sigma_8, \Gamma, n_{\rm
s}$ and $\Omega_\Lambda$.  
\end{abstract} 
 
\section{Introduction}
\label{sec:intro}

For a thorough introduction to cosmic shear, see the review of
R\'efr\'egier \cite{Ref03} in this volume.

Since its first detection in 2000, cosmic shear has proven to be a valuable
tool for cosmology. Constraints on cosmological parameters, in particular the
(dark+luminous) matter density $\Omegam$ and the power spectrum
normalization $\sigma_8$ has been obtained from cosmic shear surveys with
observed areas of up to several dozen square degrees (see Table
\ref{tab:surveys}).

\begin{table}[b]
\begin{center}
\caption{Recent and ongoing cosmic shear surveys. $m_{\rm lim}$ is the limiting
magnitude and $z_0$ is the redshift parameter of the source galaxy distribution,
see also eq.\ (\ref{prob-def}).}
\label{tab:surveys}
\begin{tabular}{llllll} \hline\hline
work & instrument & area & $m_{\rm lim}$ & $z_0$ & Ref. \\ \hline
Hoekstra et al.\ (2002), RCS & CFHT, CTIO & 53 $\Box^2$ & $R=24$
 & 0.55 & \cite{H02} \\
Jarvis et al.\ (2003)        & CTIO  & 75 $\Box^2$ & $R=23$ & 0.66 & \cite{J03} \\
van Waerbeke et al.\ (2002), Virmos-Descart & CFHT  & 6.5 $\Box^2$ & $I=24$ & $
\sim$ 1 & \cite{vW02} \\
R\'efr\'egier et al.\ (2002), HST MDS       & WFPC2 & 0.36 $\Box^2$ & $I=23.5$
& 0.9 & \cite{R02} \\
Maoli et al. (2001) & VLT FORS1 & 0.64 $\Box^2$ & $I=24
$ & 0.8 & \cite{M02} \\ \hline\hline
\end{tabular}
\end{center}
\end{table}

A cosmic shear survey has to cover a large area containing hundreds of
thousands of galaxies, whose shapes can be determined. Because
telescope time is limited, one has to carefully choose the locations
of the pointings, in other words, the geometry of the survey: On the
one hand, many independent lines-of-sight are preferable, lowering the
sampling or ``cosmic variance''. On the other hand, it is important to
measure the shear on a large range of angular scales.  Even with
modern wide-field imaging cameras, separations of more than $\sim$ 1
degree cannot be accessed by individual fields-of-view --- one has to
observe some fields near to each other and measure galaxy shape
correlations across fields.  Thus, a tradeoff between
``clustering'' and wide separation of pointings has to be
found. In this contribution, we present a method to compare different
survey configurations with respect to their ability to constrain
cosmological parameters. For a more detailed description, see \cite{PaperII}.

The survey geometries considered here consist of $n$ images which are placed
into $P$ patches of radius $R$, each patch containing $N$ pointings (thus $n = P
\cdot N$). The individual patches are assumed to be uncorrelated i.e.\ widely
separated and randomly placed on the sky. The largest scale on which cosmic
shear can be probed in these cases is $2 R$. In addition to these patch
geometries, a survey is considered where all $n$ individual fields are
uncorrelated. For a given $n$, this survey has the smallest possible cosmic
variance, but the largest scale to probe cosmic shear is only $\sqrt 2$ times
the image size.

We use a total image number of $n = 300$. The individual images are $13^\prime
\times 13^\prime$-fields (which is roughly the field-of-view of VIMOS).
This corresponds to a survey area of about 14 square degrees. The
number of galaxies, for which a shape measurement is feasible, is set to
30 per square arc minute, corresponding to a limiting $R$-band
magnitude of about 25.5.


\section{Cosmological model} 
\label{sec:cosm}

For the power spectrum of the matter fluctuations, we assume an initial
power law $P_{\rm i} \propto k^{n_{\rm s}}$, the transfer function for Cold
Dark Matter from \cite{BBKS} and the fitting formula for
the non-linear evolution of \cite{PD}. The redshift
distribution of the source galaxies is set to \cite{Smail}
\begin{equation}
p(z) \dd z = \frac{\beta}{z_0 \Gamma\left(3/\beta\right)} \left(
\frac{z}{z_0} \right)^2
\ee^{- \left( z/z_0 \right)^\beta} \dd z,
\label{prob-def}
\end{equation}
where $\Gamma$ denotes the Eulerian gamma function.
Our reference cosmology is a flat $\Lambda$CDM model with $\Omegam = 0.3$,
$\Omega_\Lambda = 0.7$, $n_{\rm s} = 1$, the shape parameter $\Gamma=0.21$ and the normalization
$\sigma_8=1$. The parameters of the
redshift distribution are $z_0=1$ and $\beta=1.5$, which corresponds to a mean
source redshift of $\approx 1.5$.

\section{Covariance of two-point statistics of cosmic shear}


We use three different second order statistics of cosmic shear for our analysis,
the two two-point correlation functions $\xi_\pm$ and the aperture mass
$\average{\Mapsq}$.

\begin{figure}[t!]
\vspace*{1.25cm}
\begin{minipage}[c]{\textwidthhalf}
\includegraphics[scale=0.60]{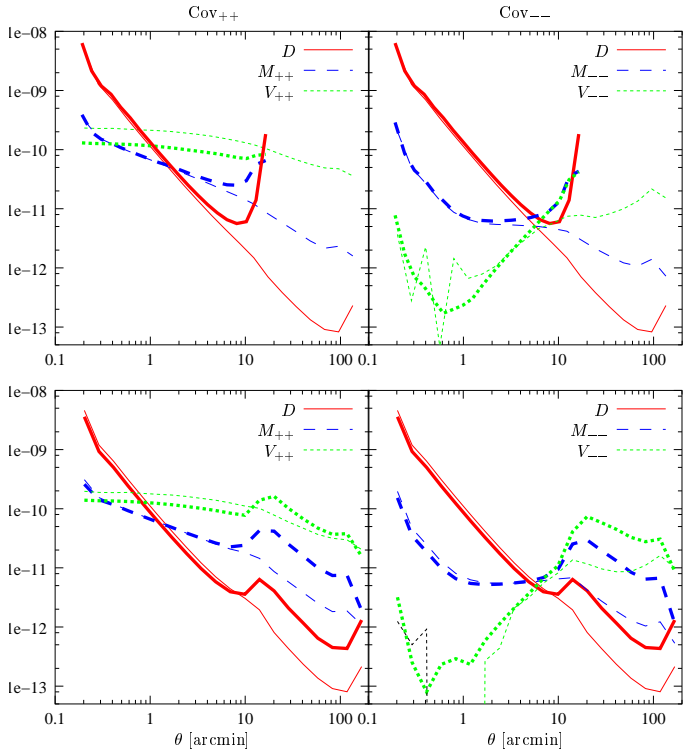}
\end{minipage}
\begin{minipage}[c]{\textwidthrhalf}
\caption{The diagonal elements of the covariance
matrices (\ref{Cov}), split up into the individual terms $D$, $M$ and
$V$.
\emph{Upper row:} 300 single uncorrelated images, where
the largest scale is $\sqrt{2} \cdot 13^\prime$ (thick lines)
and a patch geometry with $N=60$ and $R=80^\prime$ (thin
lines). \emph{Lower row:} $N=10$, $R=100^\prime$ (thick lines) and $N=60$,
$R=100^\prime$ (thin lines).
}
\label{fig:diag}
\end{minipage}
\end{figure}

The covariance matrices of the two-point correlation functions are defined as
follows:
\begin{equation}
{\rm Cov}_{\pm \,\pm} \equiv {\rm Cov}(\hat\xi_\pm,\theta_i;\hat\xi_\pm,\theta_j)
=\average{\left(\hat \xi_\pm(\theta_i)-\xi_\pm(\theta_i)\right)\left(\hat
\xi_\pm(\theta_j)-\xi_\pm(\theta_j)\right)},
\label{Cov}
\end{equation}
where $\hat \xi_\pm$ are unbiased estimators of $\xi_\pm$.
The covariance of the aperture mass is defined in complete analogy.

The covariance matrices of the correlation functions consist of three different
terms: a shot noise term ($D$) which appears only on the diagonal and depends solely
on the intrinsic ellipticity dispersion, a term which depends on cosmic
variance only ($V$) and a mixed term ($M$). In Fig.\ \ref{fig:diag}, the
diagonal elements of the covariance matrices are plotted, split up into the
individual terms. For $\, \Cov_{++}$, the cosmic variance $V$ dominates on
scales larger than about one arc minute, whereas $\Cov_{--}$ is dominated by
$D$. As shown in the upper row of Fig.\ \ref{fig:diag}, there is about a factor
of two in $V_{++}$ between the two extreme geometries
regarding cosmic variance. The lower row of panels of Fig.\ \ref{fig:diag}
compares patch geometries with the same radius, but with a low ($N=10$) and a
high ($N=60$) image density in the patches. In the first case, there is a sharp
transition at the image boundary scale, which is not present for the denser
patches.

\section{Analysis}

\subsection{Maximum likelihood}

Using the covariance matrices, we construct a figure-of-merit
\begin{equation}
\chi^2(p) \equiv \sum_{ij} \left( \xi_i(p) - \xi^{\rm t}_i \right)
\Cov_{ij}^{-1} \left( \xi_j(p) - \xi^{\rm t}_j \right), \;\;\;\;\;\;\; \xi_i
\equiv \xi(\theta_i),
\label{chi-sqr}
\end{equation}
where the superscript $^{\rm t}$ denotes the fiducial model (see Sect.\
\ref{sec:cosm}) and $p$ is a set of cosmological parameters which is tested
against this model. This figure-of-merit can be calculated using the two
correlation functions $\xi_+$ and $\xi_-$ separately, or the combination of
both; the resulting functions are called $\chi^2_+$, $\chi^2_-$ and 
$\chi^2_{\rm tot}$, respectively.

We calculated (\ref{chi-sqr}) for a number of different survey
geometries. The most interesting constraints on cosmological
parameters from cosmic shear are those on $\Omegam$ and $\sigma_8$. We
kept all other parameters fixed and reparametrized $\sigma_8$ to
$\Sigma_8 \equiv \sigma_8 / [ 0.41 + 0.59 \left(
\Omegam/0.3 \right)^{-0.68} ]$ in order to compensate for the
high elongation of the $\Omegam$-$\sigma_8$- likelihood contours.

\begin{figure}
\vspace*{1.25cm}
\begin{minipage}[c]{\textwidthhalf}
\epsfig{figure=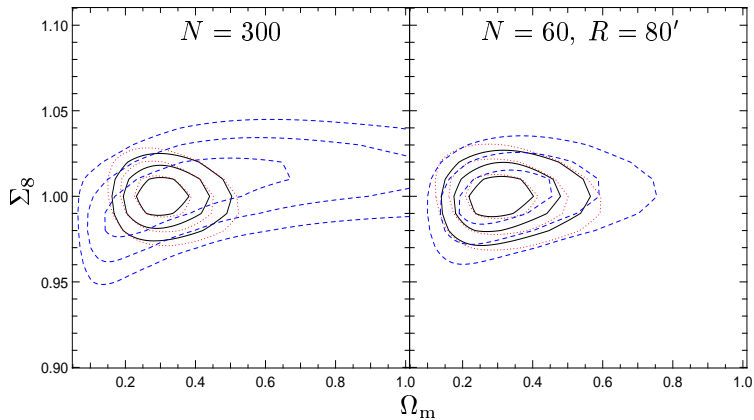,width=10cm}
\end{minipage}
\begin{minipage}[c]{\textwidthrhalf}
\caption{1-$\sigma$, 2-$\sigma$ and 3-$\sigma$ confidence contours of the
figure-of-merit (\ref{chi-sqr}). The parameter $\Sigma_8$ is
proportional to $\sigma_8$, see text. Solid lines are from the combination of
$\xi_+$ and $\xi_-$, dotted (dashed) lines are calculated using $\xi_+$ ($\xi_-$) alone. The left
panel shows the results for the 300 uncorrelated images, the right panel
is for a $(N=60, R=80^\prime)$-patch geometry.}
\label{fig:contour1}
\end{minipage}
\end{figure}

In Fig.\ \ref{fig:contour1}, the $\Omegam$-$\Sigma_8$-likelihood contours
of (\ref{chi-sqr}) are plotted, for two extreme geometrical configurations
regarding cosmic variance: the uncorrelated images and the $(N=60,
R=80^\prime)$-patch geometry. 
Clearly, the $\chi^2_-$-contours are more
extended in the case of the uncorrelated images than for the patch
geometry. This is because $\xi_-$ contains much information on large
scales which is absent in the case of the uncorrelated images. In contrast,
the $\chi^2_+$-contours are tighter in this case than
for the patch geometry.

Furthermore, in both cases, the difference between $\chi^2_+$ and
$\chi^2_\tot$ are small. Thus, most of the information concerning
these two cosmological parameters is contained in $\xi_+$; the
additional information coming from $\xi_-$ is relatively small.

\begin{figure}
\vspace*{1.25cm}
\begin{center}
\epsfig{figure=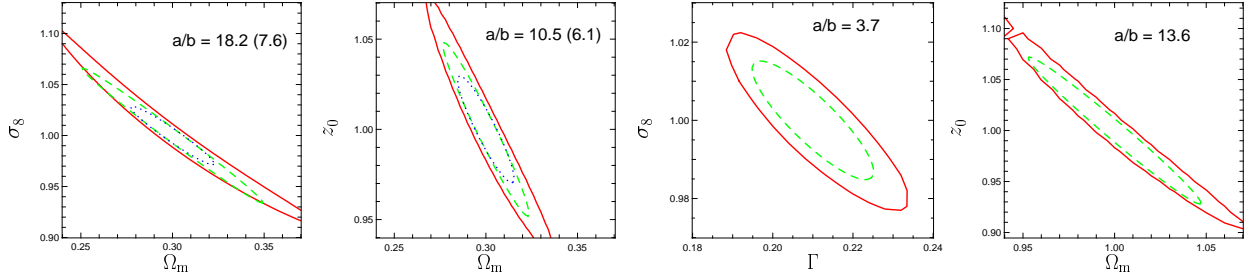,width=16.5cm}
\end{center}
\vspace*{0.25cm}
\caption{1-$\sigma$-likelihood contours (solid lines) compared with the
1-$\sigma$-error ellipse from the Fisher matrix: the dashed ellipse is for
a flat Universe (as it is the case for the likelihood contours), the dotted
one is for $\Omega_\Lambda = 0.7$. a/b is the axis ratio of the ellipses
(the case $\Omega_\Lambda = 0.7$ is in parentheses). The configuration is a
survey of 300 uncorrelated images, we used the combined $\xi_+$ and $\xi_-$.
\label{fig:comb2}
}
\end{figure}

\subsection{Fisher information matrix}

The Fisher information matrix $F$ is defined as the expectation value of
the second derivative of the likelihood function. From the Cram\'er-Rao
inequality, we get a minimum variance bound (MVB)
$\sigma(p_i) = \sqrt{(F^{-1})_{ii}}$
for any unbiased estimator of a parameter $p_i$.

In Fig.\ \ref{fig:comb2}, we compare the MVB from the Fisher matrix
with the 1-$\sigma$-contours of the likelihood, using $\xi_+$ in both
cases. As expected, the likelihood contours are larger than the
1-$\sigma$-ellipse from the Fisher matrix. The orientation of the
Fisher ellipse coincides with the likelihood shapes, i.e.\ the
direction of the minimal and maximal degeneracy of parameters is
recovered. The larger the degeneracy between two parameters, the
larger is the deviation between the local approximation by the Fisher
matrix and the likelihood function.

Next, we assume that the parameters $\Omegam, \sigma_8, \Gamma$ and $n_{\rm s}$
are to be determined simultaneously from the survey and calculate the MVB for
these parameters, using different survey geometries. We now perform the analyses
using the $\Map$-statistics which is thought to be the most useful statistics
for cosmic shear because of its ability to separate E- from B-modes.

\begin{figure}[ht!]
\vspace*{1.25cm}
\begin{center}
\epsfig{figure=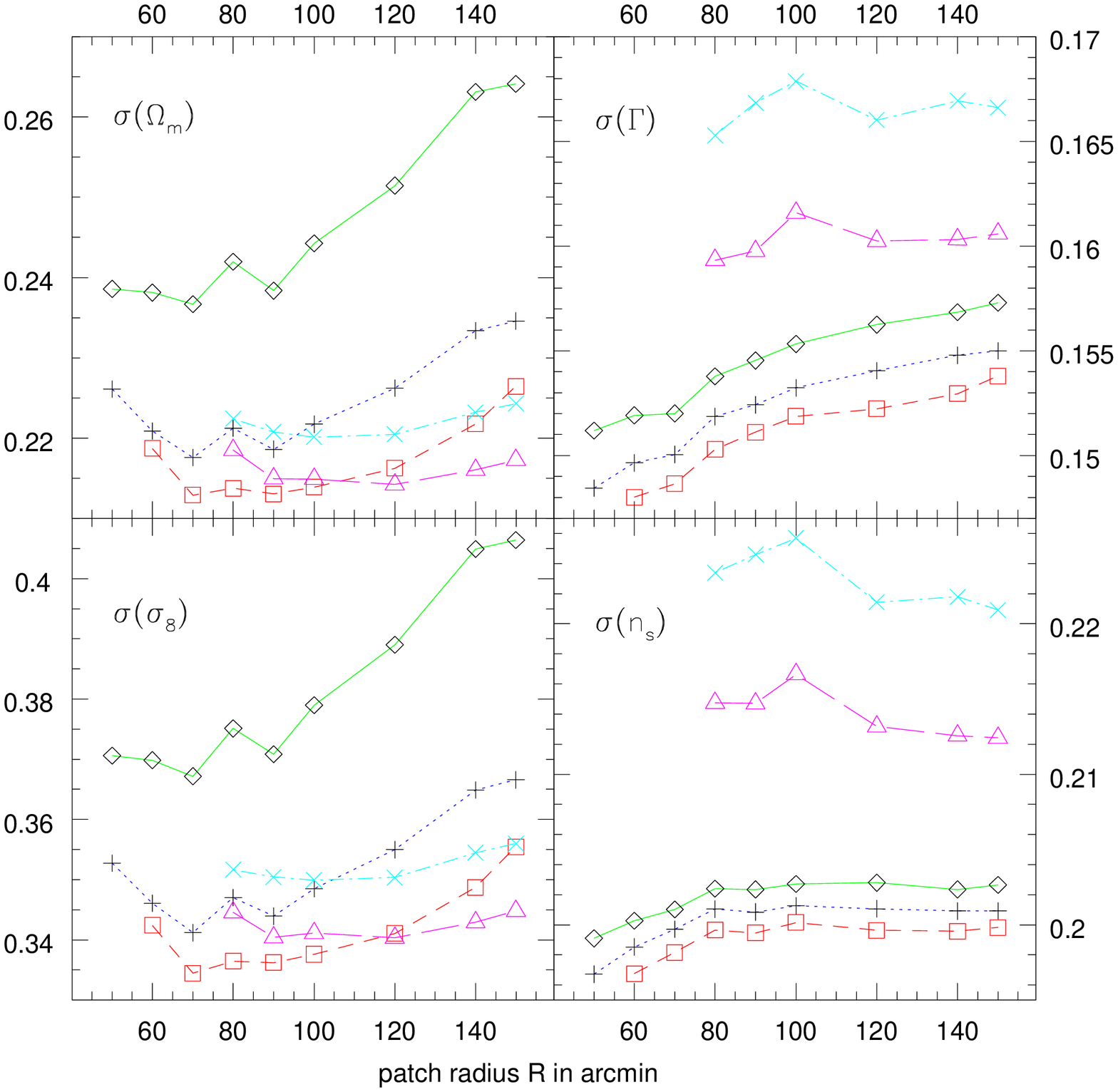,width=9.7cm}
\epsfig{figure=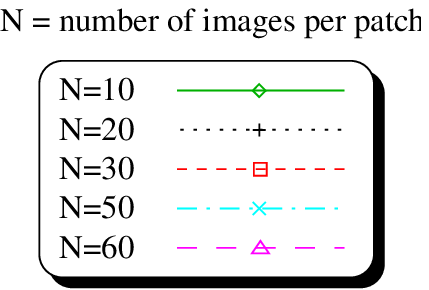,width=3.9cm}
\end{center}
\vspace*{0.25cm}
\caption{The MVB for the parameters $\Omegam, \Gamma, \sigma_8$ and $n_{\rm
s}$, using the
$\average{\Mapsq}$-statistics. 
$\Omegam, \Gamma, \sigma_8$ and $n_{\rm s}$ are assumed
to be determined from the data, all other parameters are kept fixed. Each
point in the plot represents a patch geometry with $N$ images in patches of
radius
$R$.}
\label{fig:mvb}
\end{figure}

The MVB for these parameters for a variety of survey geometries is plotted in
Fig.\ \ref{fig:mvb}. The smallest values for the MBVs are obtained for a
configuration with $N=30$ and $R=60^\prime-70^\prime$. Note
the similarity of the curves for $\Omegam$ and $\sigma_8$ which both determine
more or less the amplitude of the power spectrum, and those for $\Gamma$ and
$n_{\rm s}$ which are responsible for the shape.

\section{Conclusions}

The best constraints on cosmological parametes are obtained for a survey with
$N=30$ images distributed in patches on the sky. The dependence on the patch
radius is quite small. The MVBs of some combinations of cosmological parameters
to be determined from a cosmic shear survey with $N=30$ and $R=60^\prime$ are
given in Table \ref{tab:mvb}. The cosmological constant is only poorly
constrained. For $\Omegam, \sigma_8, \Gamma$ and $n_{\rm s}$, the optimal survey
consists of small patches of about $R=60$ arc minutes of radius, with $N=30$
images randomly distributed in each patch. The difference of the MVB makes up to
25 percent for different survey geometries, thus a 25 percent improvement on the
determination on cosmological parameters can be obtained solely by choosing the
appropriate survey geometry.

\begin{table}[b]
\begin{minipage}[c]{\textwidthhalf}
\caption{Each row shows the MVBs for a combination of three or four cosmological
parameters which are to be determined simultanously from a cosmic
shear survey.}
\label{tab:mvb}
\end{minipage}
\begin{minipage}[c]{\textwidthrhalf}
\begin{tabular}{lllll}\hline\hline
\rule[-2mm]{0mm}{6mm} $\Omegam$ & $\sigma_8$ & $\Gamma$ & $\Omega_\Lambda$ & $n_{\rm s}$\\
\hline
0.18 & 0.28 & 0.04 & & \\
0.20 & 0.28 & 0.10 & 0.41 & \\
0.22 & 0.34 & 0.15 & & 0.20 \\ \hline\hline
\end{tabular}
\end{minipage}

\end{table}

\acknowledgements{We thank Peter Schneider for useful discussions and helpful comments on the manu\-script.
}

\bibliographystyle{aa}

\vfill 
\end{document}